# From Hume to Jaynes: Induction as the Logic of Plausible Reasoning

Tommaso Costa[1,2,3]

[1] GCS-fMRI, Koelliker Hospital and Department of Psychology, University of Turin, Turin, Italy.
[2] FOCUS Laboratory, Department of Psychology, University of Turin, Turin, Italy.
[3] Neuroscience Institute of Turin (NIT), Turin, Italy.

e-mail address: tommaso.costa@unito.it

# Abstract

The problem of induction — how experience can justify belief — has persisted since Hume (Hume, 1748) exposed the logical gap between repeated observation and universal inference.

Traditional attempts to resolve it have oscillated between two extremes: the probabilistic optimism of Laplace (Laplace, 1814/1951) and Jeffreys, who sought to quantify belief through probability, and the critical skepticism of Popper, who replaced confirmation with falsification. Both approaches, however, assume that induction must deliver certainty or its negation.

In this paper, I argue that the problem of induction dissolves when recast in terms of logical coherence (understood as internal consistency of credences under updating) rather than truth. Following E. T. Jaynes, probability is interpreted not as frequency or decision rule but as the extension of deductive logic to incomplete information. Under this interpretation, Bayes's theorem is not an empirical statement but a consistency condition that constrains rational belief updating. Induction thus emerges as the special case of deductive reasoning applied to uncertain premises.

Falsification appears as the limiting form of Bayesian updating when new data drive posterior plausibility toward zero, while the Bayes Factor quantifies the continuous spectrum of evidential strength.

Through analytical examples — including Laplace (Laplace, 1814/1951)'s sunrise problem, Jeffreys (Jeffreys, 1939)'s mixed prior, and confidence-based reformulations — I show that only the logic of plausible reasoning unifies these perspectives without contradiction. Induction, properly understood, is not the leap from past to future but the discipline of maintaining coherence between evidence, belief, and information.

# 1. Introduction

Since the dawn of thought, human beings have tried to turn experience into knowledge. We observe regularities in the world and instinctively generalize them into laws: the Sun has always risen; therefore, it will rise again tomorrow; all ravens we have seen are black; therefore, all ravens are black. This way of reasoning — moving from particular instances to general propositions — is what we call induction. Yet the rational justification of this process has troubled philosophers for centuries.

As early as the Pyrrhonian skeptics, thinkers like Sextus Empiricus had noticed that no finite number of observations can logically establish a universal statement. The problem, restated in its modern form by David Hume (Hume, 1748) in the eighteenth century, is that there is no logical reason why the future should resemble the past. Why should repeated success make us rational in expecting another success? Science, whose very method depends on generalization, seems to rest on a fragile foundation.

With Bayes and Laplace, a new hope appeared: perhaps probability could provide a logical bridge from the known to the unknown. If certainty is impossible, we can at least assign degrees of belief to hypotheses and revise them as evidence accumulates. This was the birth of Bayesian reasoning — knowledge as the coherent revision of belief in light of new data. Yet Laplace (Laplace, 1814/1951)'s famous sunrise problem soon revealed a deep limitation. Even after millions of successful sunrises, the probability that "the Sun will rise forever" remains exactly zero. Probabilistic reasoning works flawlessly as a calculus, but it does not answer Hume (Hume, 1748)'s challenge.

In the twentieth century, Harold Jeffreys (Jeffreys, 1939) attempted to repair this deficiency by introducing priors with point masses at special values, while R.A. Fisher (Fisher, 1935) sought an alternative with his fiducial inference, later reformulated by others as confidence or extended likelihood. The idea is intuitively appealing: instead of asking for the probability that a hypothesis is true, we ask how strongly the data warrant confidence in it. After a long sequence of confirming observations, one may "accept" a hypothesis with full confidence and withdraw that acceptance as soon as contradictory evidence appears. But this move does not solve the induction problem — it merely replaces a question of justification with a rule of decision. It changes the vocabulary, not the logic.

In this paper, I propose to look at induction from a different standpoint, one inspired by E. T. Jaynes (Jaynes, 2003). Induction is not a leap of faith from past to future; it is an extension of logic to cases of incomplete information. In this view, probability is not a property of the world but a measure of plausibility reflecting our state of knowledge. The goal is not to prove that nature is uniform, but to reason consistently with the information we have — to assign beliefs and make decisions that are coherent, revisable, and explicitly conditioned on evidence.

Under this interpretation, the "problem of induction" is not a flaw in nature but a question of rational coherence. We do not need to prove that the Sun will rise forever; we need only show that, given what we know, believing it will rise tomorrow is the most consistent and informative stance. Probability, likelihood, and confidence each play a role in this process: probability orders our beliefs; likelihood measures the agreement between model and data; confidence calibrates decisions. None of them, taken alone, resolves induction — but together they clarify what it truly means.

In what follows, I examine why confidence-based approaches fail to overcome Hume (Hume, 1748)'s paradox and develop an alternative framework grounded in the logic of plausible reasoning. I argue that the rational content of induction lies not in claiming certainty, but in maintaining coherence under uncertainty — in our ability to update beliefs, to act upon them, and to revise them when evidence changes. Induction, in this sense, is not faith in the future, but the logic of rational consistency.

## 1.1 Overview of the Argument

The structure of the argument can be summarized as a logical trajectory from *skepticism* to *coherence*.
The paper proceeds through five conceptual transitions:

1. Hume – The Skeptical Challenge: shows that no finite observation can justify universal inference; the uniformity of nature cannot be proven logically.
2. Laplace – The Probabilistic Bridge: transforms Hume's puzzle into a quantitative problem, proposing probability as a measure of rational expectation, yet revealing *probability dilution*: even perfect evidence cannot confirm a universal law.
3. Jeffreys – The Objective Prior: attempts to restore universality by assigning prior mass to immutable laws, solving the arithmetic but not the logic of induction.
4. Popper and Fisher – The Procedural Turn: replace belief with decision; falsification and confidence provide rules for action, not coherent justification.
5. Jaynes – The Logical Resolution: reframes probability as the *extension of logic to incomplete information*, unifying induction, falsification, and maximum entropy under the single principle of coherence.

This sequence defines the central thesis of the paper: induction is not a leap from past to future but the maintenance of coherence between evidence, belief, and information. Figure 1 below illustrates the conceptual flow from historical paradox to logical unification.

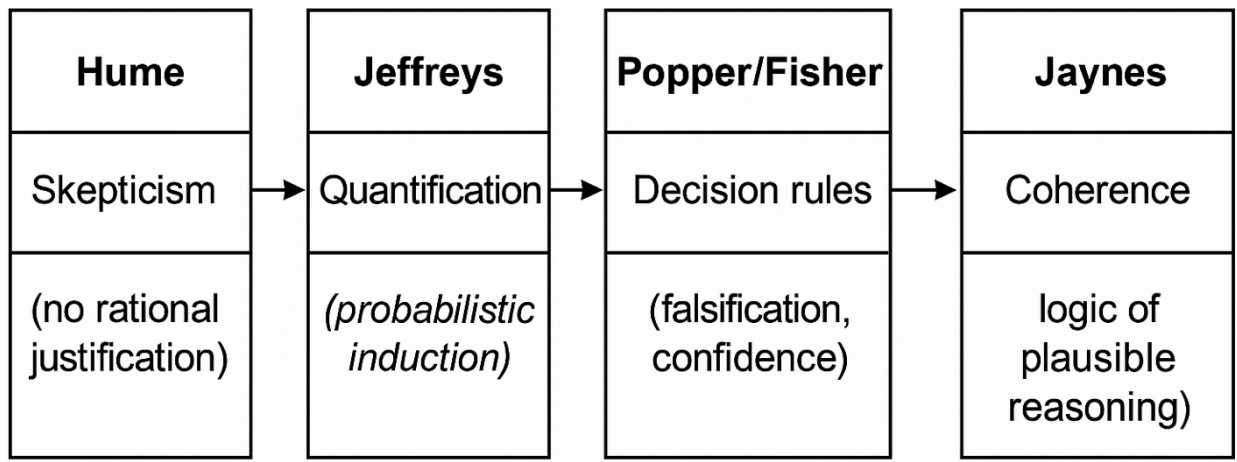

**Figure 1.** Conceptual progression from Hume's problem of induction to Jaynes's logic of plausible reasoning. Each step reframes the relationship between evidence and belief: Hume exposes the logical gap, Laplace quantifies uncertainty, Jeffreys seeks objectivity through priors, Popper and Fisher define procedural rules for action, and Jaynes unifies them through the principle of coherence.

## 2. Background: From Hume to Jaynes

The question of how we can rationally move from past observations to future expectations — the classical problem of induction — has shaped philosophical and scientific thought for centuries. It was David Hume (Hume, 1748) who expressed the problem most clearly: the assumption that the future will resemble the past can be justified neither by experience, which already presupposes it, nor by logic, since no contradiction arises in imagining a world where regularities suddenly break. Hume concluded that our faith in the uniformity of nature is not rationally grounded but the product of psychological habit. Science, then, seems to rest on a kind of well-disciplined optimism rather than logical certainty.

A century later, Pierre-Simon Laplace (Laplace, 1814/1951) sought to turn Hume (Hume, 1748)'s paradox into a calculation. If certainty is impossible, perhaps uncertainty can at least be measured.

He adopted what we now call Bayes's theorem as a rule for updating beliefs in light of evidence:

$$P(G \mid E) = \frac{P(E \mid G)\,P(G)}{P(E)}$$

Laplace (Laplace, 1814/1951) applied this rule to the famous *sunrise problem*. Let θ represent the true but unknown frequency of sunrises. Assuming a uniform prior on the interval $0 < \theta < 1$ and having observed $n$ consecutive successful sunrises, the posterior distribution becomes

$$f(\theta \mid T_n = n) = (n+1)\,\theta^n, \quad 0 < \theta < 1$$

From this, the predictive probability that the Sun will rise again tomorrow is

$$P(E_{n+1} = 1 \mid T_n = n) = \frac{n+1}{n+2}$$

This probability approaches 1 as $n$ grows, apparently confirming our expectation that repeated success increases belief. Yet the probability of the universal hypothesis — "the Sun will rise *forever*," corresponding to $\theta = 1$ — remains exactly zero, because the point $\theta = 1$ lies outside the continuous space $(0,1)$. No finite amount of evidence can ever justify a universal statement. This limitation, later called *probability*

*dilution*, reveals that probabilistic induction can predict the next event but never confirm a law.

In the early twentieth century, Harold Jeffreys (Jeffreys, 1939) tried to overcome this by modifying the prior. In his Theory of Probability (1939),, he proposed assigning part of the prior mass directly to the point $\theta = 1$:

$$\pi(\theta) = \frac{1}{2}\delta(\theta - 1) + \frac{1}{2}U(0,1)$$

where $\delta(\theta - 1)$ is a Dirac delta at the boundary. Under this mixed prior, the posterior probability that the Sun will rise forever becomes

$$P(\theta = 1 \mid T_n = n) = \frac{n + 1}{n + 2}$$

which indeed tends toward one as $n$ increases, though it never reaches it for finite $n$. The maneuver solves the arithmetic but not the logic: the choice of the prior's point mass remains arbitrary and cannot be inferred from data. The problem of induction is therefore only displaced, not resolved.

Karl Popper accepted Hume's (Hume, 1748) skepticism and recast it as a principle of scientific method. Science, he argued, advances not by confirming laws but by falsifying them. Assigning probability one to any hypothesis $G$ is irrational, because it makes learning impossible:

$$P(G) = 1 \Rightarrow P(G \mid E) = 1 \text{ for all } E$$

Once belief reaches certainty, no new observation can change it. For Popper, rationality consists in keeping hypotheses perpetually open to refutation. This view preserves the critical spirit of science but offers no quantitative way to compare unfalsified theories or express degrees of support — a gap that real scientific practice inevitably fills with probabilistic reasoning.

Around the same period, R. A. Fisher (Fisher, 1935) introduced *fiducial inference* as a third path between Bayes and frequentism. He sought a distribution for the parameter itself derived purely from data, without invoking priors. The resulting fiducial or confidence density can be represented as

$$c(\theta; t) = \frac{\partial\, P(T^* \geq t \mid \theta)}{\partial \theta}$$

and the corresponding confidence assigned to an interval $CI(t)$ is

$$C(\theta \in CI(t)) = \int_{CI(t)} c(\theta; t)\, d\theta = 1 - \alpha$$

This measure expresses a kind of "trust" in parameter values calibrated to frequency properties but independent of prior belief. Modern extensions — such as the *extended likelihood* or *confidence likelihood* — preserve this operational spirit: they allow one to act as if a hypothesis were true with full confidence after enough confirming evidence, and to withdraw that acceptance when contrary data appear. Yet this approach too remains procedural. It defines when to accept a hypothesis, not why we should believe it beyond the observed cases.

It was Edwin T. Jaynes (Jaynes, 2003) who finally reframed the issue in logical terms. Probability, he argued, is neither a frequency nor a subjective bet, but the extension of logic to situations of incomplete information. Its rules follow from requirements of internal consistency, not from empirical regularities. The fundamental identity,

$$P(G \mid E) = \frac{P(E \mid G)\, P(G)}{P(E)}$$

is, for Jaynes, not an empirical rule but a theorem of rational coherence. Induction, in this view, is not an assumption about nature's uniformity but a logical operation that preserves consistency among beliefs when information changes. The question "Why should the future resemble the past?" becomes "Given what we know, what is the most consistent assignment of plausibility?" Under this interpretation, the ancient problem of induction dissolves: it is no longer about proving uniformity in nature, but about reasoning coherently in the face of uncertainty.

From Hume's skepticism to Jaynes's reconstruction, the story of induction marks a profound shift — from ontology to logic, from seeking truth to maintaining coherence. Laplace made uncertainty measurable; Jeffreys sought objectivity through priors; Popper replaced confirmation with falsification; Fisher reframed inference as confidence; and Jaynes revealed that probability itself is the logic of plausible reasoning. What began as a metaphysical problem ends as a principle of rationality: the aim of induction is not certainty, but coherence under uncertainty.

## 2.1 Positioning within contemporary accounts of induction and confirmation

While the historical arc from Hume to Jaynes clarifies why "induction" should be reframed as coherent belief revision, it is useful to situate this view within contemporary accounts of inductive inference. In Bayesian confirmation theory, degrees of belief are tied to coherence requirements and to how evidence changes the plausibility of hypotheses (Howson & Urbach, 1993; Sprenger & Hartmann, 2019).

Within this tradition, Bayes's theorem is not a descriptive law but a normative constraint on rational credence; confirmation is graded, comparative, and essentially contrastive across rival models. This perspective resonates with Jaynes's program, though it emphasizes the epistemology of confirmation (what it is for evidence to confirm a hypothesis) more than Jaynes's information-theoretic foundations.

A complementary line connects rational credence to epistemic accuracy: on scoring-rule grounds, coherent credences maximize expected accuracy (Joyce, 1998). Subsequent developments have strengthened this bridge between coherence and accuracy-based justification. Greaves and Wallace (2006) demonstrated that conditionalization uniquely maximizes expected epistemic utility, providing a formal vindication of the Bayesian update rule. Pettigrew (2016) extended this program by showing that the laws of probability can be derived from accuracy-dominance principles, thus interpreting coherence not merely as internal consistency but as the rational strategy for "getting things right." Leitgeb (2017) further integrated these ideas within a unified account of belief stability, linking probabilistic coherence with the resilience of rational belief across time. Together, these contributions establish that coherence and accuracy are not competing ideals but complementary dimensions of rationality: coherent beliefs are precisely those that remain stable and accuracy-conducive under new evidence.

Relatedly, programs in probabilistic truth-likeness and verisimilitude investigate how graded belief can track closeness to truth without collapsing into binary acceptance, providing a formal backdrop for the idea that falsification is a limit case of continuous evidential re-weighting. Formal learning theory provides a third vantage point: under broad conditions, Bayesian updating converges (in various senses) to the data-generating hypothesis or to its best available approximation in the model class, thereby addressing the pragmatic core of Hume's worry (Williamson, 2000).

Beyond its epistemic and formal virtues, the coherence-based view of Bayesian reasoning also connects with broader discussions of scientific realism. Recent analyses emphasize that coherence is not only a constraint on rational belief but also a criterion for epistemic reliability in scientific practice. As Sprenger (2021) argues, Bayesian coherence provides a realist justification for probabilistic inference: it explains why coherent updating tends, in the long run, to align credences with the structure of the world rather than with mere pragmatic convenience. In this sense, coherence is both a logical and an ontological bridge—linking rational belief revision with the realist expectation that successful theories approximate truth.

Finally, recent discussions of likelihood-based and confidence-based proposals (e.g., fiducial or extended likelihood) argue for decision-calibrated tools with strong long-run properties. Our stance is sympathetic to their operational merits but stresses a distinction: procedures that optimize error-control or coverage do not, by themselves, supply a normative theory of single-case rational belief. In what follows we therefore treat likelihood and confidence as valuable instruments for calibration and action, while reserving to Bayesian coherence the role of a general logic for belief and learning. See §6 for the relation between coherence, falsification, and accuracy-based justifications, and see §6.1 for the relation between coherence and epistemic accuracy.

## 3. Why "confidence" does not solve induction

The recent revival of confidence-based reasoning—also called extended likelihood or confidence likelihood in the sense of Lee (2025)—seeks to overcome a classical limitation of Bayesian inference: the impossibility of assigning probability 1 to a universal law. In this framework, the term *confidence* does not refer to frequentist confidence intervals, but to a reinterpretation of the likelihood function as a decision-calibrated measure of evidential support, designed to mirror the long-run properties of confidence coverage while retaining a likelihood-based formulation. Lee argue that, if probability cannot express full acceptance, *confidence* can. After sufficient consistent data, a hypothesis may be accepted with confidence equal to 1 and rejected with confidence 0 at the first contradiction. At first glance, this appears to solve the ancient puzzle of induction. In reality, however, it merely transfers the problem from the level of epistemic justification to that of procedural decision.

The approach traces its lineage to Fisher's *fiducial inference* and was reformulated by Lee as an *extended-likelihood framework*. Its central idea is to replace the probability density $f(\theta \mid t)$ with a confidence density $c(\theta; t)$, defined as:

$$c(\theta; t) = \frac{\partial P(T^* \geq t \mid \theta)}{\partial \theta}$$

Integrating this function over the observed confidence interval $CI(t)$ yields the confidence measure

$$C(\theta \in CI(t)) = \int_{CI(t)} c(\theta; t) \, d\theta = 1 - \alpha$$

This formulation preserves the property known as *confidence coverage*: under repeated sampling, the probability that the true value $\theta_0$ lies within the reported confidence interval equals the stated confidence level. Formally,

$$P(\theta_0 \in CI(T_n)) = C(\theta_0 \in CI(t))$$

This equality is elegant and operationally useful. It guarantees that, across many hypothetical repetitions of the experiment, the proportion of intervals containing the true parameter matches their nominal confidence. Yet this property is purely frequentist; it says nothing about the logical or epistemic status of the statement "$\theta$ lies in this particular interval."

When applied to Laplace's sunrise model, the confidence framework yields a mixed object that assigns all mass to the boundary point $\theta = 1$ whenever all observed cases are successes ($T_n = n$):

$$c(\theta; t) = \begin{cases} \delta(\theta - 1), & \text{if } t = n \\ \text{Beta}(t+1, n-t), & \text{if } t < n \end{cases}$$

From this, the confidence assigned to the hypothesis "the Sun will rise forever," denoted $G$, becomes:

$$C(G; n) = \begin{cases} 1, & \text{if } T_n = n \\ 0, & \text{if } T_n < n \end{cases}$$

Thus, after $n$ consecutive sunrises, we may declare complete confidence ($C = 1$) that the Sun will rise forever—until the first failure, when confidence drops instantly to 0. This on–off dynamic mirrors the Popperian notion of provisional acceptance and instant falsification.

At a pragmatic level, this behavior can be advantageous. Confidence-based methods possess strong calibration properties: their numerical levels correspond to reproducible long-run frequencies, and their *Extended Likelihood Ratio* (ELR) provides an operational scale of evidential strength even when Bayesian priors are unavailable. In applied contexts—such as industrial quality control or repeated experimental designs—these features make the framework appealing and transparent.

However, the problem at stake is not operational but epistemic. Confidence tells us *how often* a rule succeeds under repetition, not *why* it should be rational to believe its outcome in a single case. The logical relation between evidence and belief—the essence of Hume's challenge—remains unaddressed.

Indeed, the confidence measure does not express a degree of belief that $G$ is true; it encodes a decision rule: act as if $G$ were true when $C(G; E) = 1$, and abandon it when $C(G; E) = 0$. The distinction is subtle but fundamental. Bayesian probability, even with all its limitations, quantifies plausibility—a continuous relation between hypotheses and evidence—whereas confidence provides a binary operational threshold, a guide for acceptance rather than reasoning.

The contrast becomes clearer in a simple testing context. For a universal hypothesis $G: \theta = 1$ and its complement $G^C: \theta < 1$, the Bayesian comparison is given by the Bayes Factor:

$$BF(G, G^C; t) = \frac{P(T_n = t \mid \theta = 1)}{P(T_n = t \mid \theta \sim \text{Beta}(1,1))}$$

which, for the sunrise model, simplifies to:

$$BF(G, G^C; t) = \begin{cases} n + 1, & \text{if } t = n \\ 0, & \text{if } t < n \end{cases}$$

The analogous ratio in the confidence framework—the Extended Likelihood Ratio (ELR)—is defined as:

$$ELR(G, G^C; t) = \frac{C(G; t)}{C(G^C; t)}$$

producing:

$$ELR(G, G^C; t) = \begin{cases} \infty, & \text{if } t = n \\ 0, & \text{if } t < n \end{cases}$$

At first glance, this seems decisive: perfect data yield infinite support for $G$. Yet such "certainty" is purely procedural—it arises because the construction assigns all probability mass to $\theta = 1$ once $t = n$, not because inference has logically established a universal truth. The system, by design, treats the boundary case as a rule of acceptance, not as an epistemic conclusion.

In philosophical terms, the confidence approach conflates acceptance with belief. Acceptance is pragmatic—the rule to act *as if* a proposition were true. Belief, by contrast, is a coherent assignment of plausibility constrained by information. The confidence measure provides the former but not the latter. It cannot explain why our confidence that the Sun will rise tomorrow should extend beyond the observed data, nor why the same rule should apply to less regular phenomena.

The deeper reason is that confidence lacks a normative theory of inference. It offers a principled way to calibrate long-run frequencies of correct decisions, but it does not tell us how to assign rational degrees of plausibility in single cases. It is therefore silent on Hume's question: *why should the past inform the future?* What it provides instead is a well-behaved policy for deciding when to act as if a hypothesis were true, without ever establishing that it is rational to believe it.

In summary, the confidence framework succeeds where the classical frequentist failed—it provides scientists with an interpretable numerical measure of acceptance and a unified operational language—but it fails where Hume demanded an answer: it does not justify the transition from repeated observation to general belief. It replaces epistemology with policy. Jaynes's logic of plausible reasoning can thus be viewed not as a rival but as a completion of confidence reasoning: it supplies the missing normative layer that connects frequency calibration to rational belief. The problem of induction remains, though expressed in a more sophisticated syntax.

The limitations of both confidence-based and falsificationist approaches point toward a deeper unity. Each of these frameworks captures a fragment of rationality: confidence formalizes the operational need for decision under uncertainty, while falsification enforces the logical discipline of revisability. Yet both remain procedural—they describe *how* scientists should act, not *why* such actions are rational. From a Jaynesian standpoint, these methods can be understood as limit cases of the broader principle of coherence. Confidence corresponds to the pragmatic boundary where degrees of belief collapse into binary acceptance for practical purposes; falsification represents the asymptotic case where posterior plausibility approaches zero under decisive evidence. What unifies them is the same underlying logic: rational belief must evolve consistently with information. The transition from §3 to §4 thus marks a shift from operational policies to the normative foundation that renders them coherent—the logic of plausible reasoning.

## 4. Induction as the Logic of Plausible Reasoning

The failure of both probability dilution and confidence-based acceptance suggests that the problem of induction cannot be solved by modifying formulas, priors, or acceptance thresholds. The issue is not computational but conceptual. The difficulty lies in how we interpret the relationship between data, hypotheses, and rational belief. What is needed is not another mechanism for assigning numbers, but a redefinition of what those numbers *mean*.

This requirement of coherent updating implies what has been called Cromwell's Rule (Lindley, 1972; Jaynes, 2003): never assign prior probability 0 or 1 to any empirical hypothesis. The reason is simple but profound. Once a proposition is granted probability 0 or 1, Bayes's rule can no longer revise it—no evidence, however overwhelming, can alter certainty. Assigning 0 or 1 is thus not an act of reasoning but of faith. In Jaynes's logic of plausible inference, rational belief must always remain open to revision; plausibility values should occupy the continuum between these limits. In this sense, Cromwell's Rule formalizes the very spirit of scientific fallibilism: we must "think it possible that we may be mistaken."

E. T. Jaynes proposed such a redefinition. His central claim is that probability is not an empirical frequency nor a decision rule, but the extension of logic to situations where information is incomplete. Under complete knowledge, logic tells us whether a proposition is true or false. Under incomplete knowledge, probability quantifies *how strongly* the available information supports one proposition over another. The rules of probability are not arbitrary conventions but the unique system consistent with the desiderata of rational reasoning.

The guiding principle is that reasoning under uncertainty must obey the same structural constraints as deductive logic: consistency, symmetry, and transparency to new information. From these requirements, the product and sum rules of probability follow uniquely. The basic relation is the familiar rule of conditional updating:

$$P(G \mid E) = \frac{P(E \mid G)\, P(G)}{P(E)}$$

but here it is interpreted not as a statement about the world, but as a constraint on rational belief. It enforces internal coherence among propositions once evidence is specified.

In this sense, Bayesian updating is not an inductive law of nature but a consistency theorem: given our prior plausibility assignments, there is only one coherent way to revise them when new data arrive. Every other method would lead to contradictions in reasoning. Jaynes often emphasized that probability theory is "the logic of science," not a theory of random events. Its subject is not the behavior of nature but the behavior of *rational thought* about nature.

This view dissolves Hume's paradox. The question "why should the future resemble the past?" is misplaced. The Bayesian does not assert that the world is uniform; they merely condition their expectations on the information available. If tomorrow's sunrise fails, beliefs will change automatically by the same rule that once supported them. The logic of plausible reasoning requires no metaphysical assumption about the world's regularity — only a commitment to coherent updating.

To formalize this, Jaynes proposed a few general desiderata for any system of reasoning under uncertainty:

1. Representation: A real number $P(A \mid B)$ represents the plausibility of proposition $A$ given $B$.
2. Consistency: If a conclusion can be reached in more than one way, every valid path must lead to the same result.
3. Qualitative correspondence with logic: When information becomes complete, probabilities reduce to truth values 0 or 1.

From these, the sum and product rules follow. The sum rule expresses how plausibility combines for mutually exclusive propositions:

$$P(A + B \mid C) = P(A \mid C) + P(B \mid C) - P(AB \mid C)$$

and the product rule governs conditional inference:

$$P(AB \mid C) = P(A \mid C)\, P(B \mid AC)$$

Together, they form the algebra of plausible reasoning. All valid methods of inference—Bayesian or otherwise—must be consistent with them.

A key insight of Jaynes's approach is that induction is not a separate kind of reasoning, but a special case of deduction applied to uncertain premises. In deductive logic, we write

$$I(G) = 1 \text{ or } 0$$

depending on whether the proposition $G$ is true or false. In plausible reasoning, we instead assign a degree of plausibility:

$$0 < P(G \mid E) < 1$$

The transition from certainty to uncertainty does not require new principles, only an extension of logic to a continuum of plausibility values.

This reinterpretation resolves the logical tension that plagued earlier formulations. Laplace's paradox arose because probability was treated as a physical property of the world; Jaynes removes this assumption. Probability lives in the mind of the reasoner, not in the world itself. It encodes information, not randomness. When evidence accumulates, the probability $P(G \mid E)$ approaches one, not because nature becomes more uniform, but because the information supporting $G$ becomes more complete.

Moreover, this framework restores a clear distinction between belief, decision, and truth. Truth ($I(G)$) is ontological; belief ($P(G \mid E)$) is epistemic; decision is pragmatic. The Bayesian logic of plausible reasoning pertains to belief: how a rational agent should assign and revise plausibilities given information. Once beliefs are assigned, decisions can be made according to a utility criterion, but that is a separate step. Induction, in this sense, is not a method for discovering truth, but for maintaining coherence in belief revision.

This interpretation also clarifies the relationship between Bayes's rule and falsification. Popper was right to insist that hypotheses must remain open to refutation; Jaynes's logic ensures this automatically. If new data $E'$ contradict the predictions of $G$, the likelihood $P(E' \mid G)$ becomes small, and therefore $P(G \mid E, E')$ decreases accordingly. The Bayesian rule accomplishes Popperian falsification smoothly and quantitatively:

$$P(G \mid E, E') \propto P(E' \mid G)\, P(G \mid E)$$

No special rule for rejection is needed: inconsistency with data leads to lower plausibility by logical necessity.

Finally, Jaynes's framework connects naturally with principles of information theory. When no specific prior information is available, the maximum entropy principle prescribes choosing the distribution that makes the fewest unwarranted assumptions while satisfying known constraints. In this way, probability becomes the language of *honest ignorance* — expressing what is known without asserting what is not. This principle completes the logical structure of induction: new data reduce uncertainty, old priors encode background knowledge, and the updating rule guarantees consistency throughout.

In summary, Jaynes's logic of plausible reasoning reframes induction not as a metaphysical inference from past to future, but as a rule of coherent reasoning under uncertainty. The ancient demand for certainty is replaced by a higher standard: internal coherence and openness to revision. Under this view, the rational scientist does not "believe" that the Sun will always rise; rather, they assign it a plausibility close to one, subject to revision should contrary evidence arise. Induction thus ceases to be a mystery and becomes a manifestation of the same logic that governs all rational thought — the logic of plausible inference.

# 5. Case Studies: The Sunrise Problem and Beyond

The abstract discussion of induction becomes clearer when translated into concrete examples. Laplace's *sunrise problem* remains the paradigmatic case because it encapsulates the logic, the appeal, and the limitations of probabilistic inference. In this section, we revisit it under three lenses — Laplace's Bayesian formulation, Jeffreys's modification, and the confidence-based reinterpretation — and then show how a Jaynesian perspective unifies them conceptually.

Consider a sequence of observations $E = \{E_1, E_2, \ldots, E_n\}$, where each $E_i = 1$ if the Sun rises on day $i$. Let $\theta$ denote the true probability that the Sun rises on a given day. The likelihood of observing $n$ consecutive sunrises is

$$L(\theta; T_n = n) = \theta^n$$

## 5.1 Laplace's Bayesian model

Laplace assumed complete ignorance about $\theta$, represented by a uniform prior on the interval $0 < \theta < 1$. By Bayes's rule, the posterior becomes

$$f(\theta \mid T_n = n) = \frac{L(\theta; T_n = n)\, P(\theta)}{\int_0^1 L(\theta'; T_n = n)\, P(\theta')\, d\theta'}$$

Substituting $P(\theta) = 1$ and $L(\theta; T_n = n) = \theta^n$ yields

$$f(\theta \mid T_n = n) = (n+1)\, \theta^n,\, 0 < \theta < 1$$

The predictive probability that the Sun will rise again tomorrow is therefore

$$P(E_{n+1} = 1 \mid T_n = n) = \int_0^1 \theta\, f(\theta \mid T_n = n)\, d\theta = \frac{n+1}{n+2}$$

After 10 000 consecutive sunrises, this gives a predictive probability of 0.9999 — very high, but still less than one. The probability that the Sun will rise **forever**, corresponding to $\theta = 1$, remains exactly zero:

$$P(\theta = 1 \mid T_n = n) = 0$$

This illustrates *probability dilution*: finite data can strengthen belief in the next event but never justify a universal law.

## 5.2 Jeffreys's mixed prior

Jeffreys introduced a prior containing a discrete point mass at $\theta = 1$ to represent the possibility of an immutable law. The prior is

$$\pi(\theta) = \frac{1}{2}\delta(\theta - 1) + \frac{1}{2}U(0,1)$$

where $\delta(\theta - 1)$ is a Dirac delta at the boundary. The posterior probability that $\theta = 1$ after observing $n$ successes become

$$P(\theta = 1 \mid T_n = n) = \frac{(n+1)/(n+2)}{1 + (n+1)/(n+2)} = \frac{n+1}{n+3}$$

which tends toward one as $n \to \infty$, but never reaches it for any finite sequence. The numerical behavior seems satisfying, yet it is achieved only by assigning prior belief to the law itself — an assumption, not a deduction.

## 5.3 Confidence-based acceptance

In the confidence approach, the result is even more striking. When all observed cases are successes, $T_n = n$, the confidence density places all mass at $\theta = 1$:

$$c(\theta; t) = \begin{cases} \delta(\theta - 1), & \text{if } t = n \\ \text{Beta}(t + 1, n - t), & \text{if } t < n \end{cases}$$

Hence the "confidence" that the law holds is

$$C(G; n) = \begin{cases} 1, & \text{if } T_n = n \\ 0, & \text{if } T_n < n \end{cases}$$

At the first failure, confidence collapses from one to zero. The procedure mimics Popper's falsification principle: acceptance until refutation. Yet it provides no reason why the initial acceptance should extend beyond the observed sample; the number "1" here reflects a decision rule, not a rational belief.

## 5.4 The Jaynesian reinterpretation

Under Jaynes's framework, the same data are analyzed not as evidence for a universal truth, but as information constraining plausible hypotheses. The question "will the Sun rise tomorrow?" is expressed as the conditional probability

$$P(E_{n+1} = 1 \mid E_1, E_2, \ldots, E_n)$$

and evaluated using the same rules of consistency that govern all probabilistic reasoning. If all evidence so far supports the model $M$ ("the Sun rises with probability $\theta$"), and no contradictory information has appeared, then the rational assignment remains the one that maximizes coherence:

$$P(E_{n+1} = 1 \mid E_1, \ldots, E_n, M) = \frac{n+1}{n+2}$$

The value is near one, but not exactly one — and this is not a flaw but a feature. The number expresses our *state of information*, not a claim about nature's essence. If tomorrow the observation contradicts expectation, the update occurs automatically via Bayes's rule:

$$P(G \mid E, E') \propto P(E' \mid G)\, P(G \mid E)$$

Thus, the rule of coherence replaces the rule of faith. Certainty becomes unnecessary; what matters is the internal consistency between what is known and what is inferred.

## 5.5 Beyond the sunrise

The same logic applies to every domain of science. The discovery that all swans observed so far are white supports the hypothesis "all swans are white," but only within the limits of observation. Encountering a single black swan forces an immediate update. In Jaynes's system, this is not a failure of induction but a normal operation of rational inference: beliefs evolve as information changes. The strength of the method lies not in guaranteeing truth but in guaranteeing coherence.

From this perspective, both Laplace's probability and Fisher's confidence are special cases of a more general framework: the logic of plausible reasoning. The difference can be summarized conceptually as follows. Probability measures *plausibility*; confidence measures *decision readiness*; and truth measures *reality*. Only the first of these can coherently mediate between evidence and belief.

The *sunrise problem* thus teaches a broader lesson. Induction, properly understood, is not a means of proving universal truths but a disciplined method for revising plausibilities. When handled within the logic of Jaynes, the ancient problem of induction ceases to be a paradox and becomes a principle of rational humility: we act on what is most coherent with the evidence, always ready to update when the world surprises us.

## 6. Discussion and Conclusion – From Truth to Coherence

The long history of the induction problem reveals a persistent tension between two ideals of rationality: the quest for certainty and the demand for coherence. The first, inherited from classical logic, seeks conclusions that are absolutely true; the second, born of scientific practice, seeks beliefs that are consistent, revisable, and proportional to the available evidence. Hume's paradox arises only when these ideals are conflated—when we expect the logic of uncertainty to yield the kind of finality that belongs only to the logic of truth.

While many previous analyses have explored the Bayesian dissolution of Hume's problem (e.g., Howson & Urbach, 1993; Sprenger & Hartmann, 2019; Talbott, 2016), the present work advances a distinctive synthesis by explicitly linking three complementary principles—coherence, falsification, and maximum entropy—within a single logical framework. Coherence provides the *normative constraint* on rational belief; falsification represents its *limiting behavior* under contradictory evidence; and the maximum entropy principle supplies the *informational foundation* that ensures honesty and minimal assumption in prior specification.

In combination, these principles yield a complete logical solution to Hume's paradox: induction is not a separate mode of inference but the *continuous extension of deductive logic* constrained by coherence, calibrated by falsification, and grounded in information theory. This unification clarifies that the apparent conflict between confirmation and refutation disappears once both are seen as special cases of rational consistency under uncertainty.

Conceptually, this approach extends Jaynes's program beyond probabilistic reasoning alone, situating it at the intersection of epistemology, information theory, and the philosophy of science. It thereby offers a unified account of how scientific reasoning remains fallible yet logically disciplined—a resolution of the induction problem not through new empirical assumptions, but through the *logical necessity of coherent belief revision*.

Popper's falsificationism replaced the unattainable goal of confirmation with the more disciplined ideal of refutation. Scientific hypotheses can never be proven true, but they can be tested severely and rejected when contradicted by data. In symbolic form, falsification expresses the simple asymmetry

$$G \to \neg E \text{ and } E \Rightarrow \neg G$$

where the observation of a black swan (E) falsifies the general law "all swans are white" (G). Yet Popper's logic is binary: hypotheses survive or perish. In actual scientific reasoning, evidence seldom speaks in absolutes. Most data are probabilistic, and hypotheses compete not as true or false, but as more or less plausible.

Jaynes's *logic of plausible reasoning* generalizes this process by embedding falsification within a continuous scale of belief. When new data $D$ are inconsistent with a hypothesis $H$, the posterior plausibility decreases smoothly according to

$$P(H \mid D) = \frac{P(D \mid H) \, P(H)}{P(D)}$$

The extent of decrease depends on how improbable the data were under $H$. A single anomaly does not automatically destroy a theory; it simply reduces its plausibility in proportion to the likelihood. Thus, Bayesian updating transforms Popper's dichotomy into a quantitative falsification principle, where degrees of refutation are expressed by the Bayes Factor

$$BF_{10} = \frac{P(D \mid H_1)}{P(D \mid H_0)}$$

Values of $BF_{10} > 1$ indicate that data favor $H_1$, while values below 1 favor $H_0$. This ratio embodies the continuous measure of evidential strength that Popper's qualitative scheme lacked. Under this view, induction and falsification are not opposites, but complementary aspects of the same rule of belief revision. When evidence aligns with expectations, posterior probability increases; when it contradicts them, it decreases. Both are special cases of the same consistency principle:

$$P(H \mid D) \propto P(H)\, P(D \mid H)$$

This equation captures what Hume declared impossible—a rational, quantitative method for learning from experience. The paradox dissolves once we shift our goal from proving the future to adapting to it coherently. Deductive and inductive reasoning thus appear as points along a single continuum. Deduction corresponds to the limit of complete information, where probabilities collapse to 0 or 1:

$$P(H \mid D) = 1 \text{ or } 0$$

Induction corresponds to the general case of incomplete information, where plausibility varies continuously between those limits. Logic is recovered as the special case of probability when uncertainty vanishes:

$$\lim_{\text{uncertainty} \to 0} P(H \mid D) = \text{truth value}(H)$$

Hence, deduction is not opposed to induction—it is contained within it as the zero-uncertainty limit of the same rational calculus.

This unified view also clarifies the relation between belief and decision. Bayesian decision theory separates inference from action: beliefs are updated by Bayes's rule, while actions are chosen by maximizing expected utility

$$a^* = \arg\max_a \sum_H P(H \mid D)\, U(a, H)$$

where $U(a, H)$ denotes the utility of taking action $a$ when hypothesis $H$ is true. The confidence framework, by contrast, collapses these two levels: it turns belief into an immediate rule for action. Jaynes's logic restores the distinction—probability encodes *what it is rational to believe*, while utility determines *what it is rational to do*. At a deeper level, Bayesian inference aligns with information theory. When prior information is scarce, the maximum-entropy principle prescribes choosing the distribution that satisfies known constraints while making no additional assumptions. Given expected-value constraints $E[f_i(x)] = F_i$, the solution

$$p(x) = \frac{1}{Z} \exp\left(-\sum_i \lambda_i f_i(x)\right),$$

maximizes Shannon entropy

$$H[p] = -\sum_x p(x) \log p(x)$$

subject to those constraints, where $Z$ ensures normalization. This connection between information, entropy, and inference had already been noted by I. J. Good (1983), who introduced the concept of *weight of evidence* as a quantitative measure of how much an observation supports one hypothesis over another. This principle expresses *honest ignorance*: it assigns probabilities that are maximally non-committal beyond what is known, linking inference with the conservation of information.

Bringing these strands together, we can now reinterpret the problem of induction. Experience does not *prove* that the future will resemble the past; it merely constrains what it is rational to expect. Coherence, not certainty, becomes the measure of rationality. This appeal to coherence raises a natural question: is coherence merely a formal constraint of rationality, or does it also carry epistemic value, guiding belief toward truth? In Jaynes's framework, coherence is first and foremost a normative requirement—the condition any rational system of beliefs must satisfy to avoid internal contradiction. Yet coherence is not epistemically inert. When applied through Bayesian updating, coherent beliefs are also truth-conducive in the long run: under broad conditions of regularity, they converge toward hypotheses that best approximate the data-generating process (as shown in learning-theoretic results such as Williamson, 2000). Thus, coherence is both necessary for rationality and sufficient—in a pragmatic, asymptotic sense—for tracking truth over time. It ensures that belief revision is guided not by psychological habit but by a logic that, when repeatedly applied, aligns plausibility with empirical adequacy. Falsification is simply the limiting case where evidence drives plausibility toward zero; confirmation is the limiting case where it approaches one. Between them lies the continuous spectrum of plausible reasoning—the genuine logic of science. Although the argument developed here is theoretical, the conception of induction as coherence has concrete implications for scientific practice. In modern data analysis—particularly in Bayesian modeling and computational neuroscience—the principle of coherent updating governs how hypotheses are revised as new evidence accumulates. Hierarchical Bayesian models, for example, embody Jaynes's logic by integrating multiple sources of uncertainty—individual, group, and measurement levels—under a single rule of consistency. When models are compared via Bayes Factors or predictive performance, scientists are not merely performing statistical calculations; they are enacting the logic of plausible reasoning, ensuring that belief updates remain transparent, proportional, and reversible. In neuroimaging or biomarker research, this means that the adoption or rejection of a model is guided not by dichotomous thresholds but by the degree to which data coherently reshape prior plausibilities. Thus, the logic of coherence provides both a normative foundation for inference and a practical discipline for learning from data, linking philosophical rationality with everyday scientific reasoning.

Despite its unifying power, the Jaynesian framework is not without limitations. Its normative strength derives from internal coherence, yet practical inference depends on modeling choices that are not uniquely determined by logic. The requirement of specifying priors introduces an element of judgment: different prior structures may yield distinct posterior conclusions, especially under limited data. Likewise, the assumption of model completeness—treating one's hypothesis space as exhaustive—can only be an idealization. Real scientific reasoning must therefore combine Jaynes's logic of coherence with empirical caution: priors should be critically examined, model classes expanded or hierarchically structured, and coherence interpreted as a guiding ideal rather than an absolute guarantee. Recognizing these boundaries does not weaken the framework; it situates it where it belongs—at the interface between rational consistency and the fallible, open-ended nature of empirical science.

From Hume's skepticism to Jaynes's reconstruction, the trajectory of thought moves from *truth* to *coherence*, from *ontology* to *epistemology*. Induction, once the great riddle of philosophy, re-emerges not as a leap of faith but as a rule of consistency: the rational art of maintaining coherence between belief, evidence, and information under uncertainty.

# Appendix A – Mathematical Notes

This appendix summarizes the key formal relations underlying the logic of plausible reasoning discussed in the main text. The goal is to make explicit how Bayes's theorem, the consistency desiderata, and the maximum entropy principle arise as logical constraints on rational inference rather than as empirical assumptions.

## A.1 Derivation of Bayes's Rule from Consistency

Let $A$, $B$, and $C$ be logical propositions. The plausibility of a conjunction AB given C must satisfy two intuitive requirements:

1. Product rule: the plausibility of $AB$ given $C$ equals the plausibility of $A$ given $C$ times the plausibility of $B$ given both $A$ and $C$:

$$P(AB \mid C) = P(A \mid C) \, P(B \mid AC)$$

2. Symmetry of conjunction: the order of propositions in a conjunction should not matter:

$$P(AB \mid C) = P(BA \mid C)$$

Combining these gives

$$P(A \mid C) \, P(B \mid AC) = P(B \mid C) \, P(A \mid BC)$$

Rearranging, we obtain Bayes's rule as a direct consequence of consistency:

$$P(A \mid BC) = \frac{P(B \mid AC) \, P(A \mid C)}{P(B \mid C)}$$

This relation is not an empirical law but the only way to assign degrees of belief consistently when evidence is updated. Every self-consistent system of inference must obey this proportionality.

## A.2 Bayes Factor as a Measure of Evidential Strength

Given two competing hypotheses $G_1$ and $G_2$ and evidence $E$, the Bayes Factor quantifies how much the data shifts our relative belief:

$$BF_{12} = \frac{P(E \mid G_1)}{P(E \mid G_2)}$$

The posterior odds are obtained from the prior odds multiplied by the Bayes Factor:

$$\frac{P(G_1 \mid E)}{P(G_2 \mid E)} = BF_{12} \times \frac{P(G_1)}{P(G_2)}$$

Thus, $BF_{12}$ acts as a multiplicative update in the space of odds. On a logarithmic scale, evidence accumulates additively:

$$\log_{10} BF_{12}^{(1,2,\ldots,n)} = \sum_i \log_{10} BF_{12}^{(i)}$$

This cumulative property explains why Bayesian learning naturally integrates sequential evidence without contradiction.

## A.3 The Principle of Maximum Entropy

When only partial information is available, the probability distribution that best represents our state of knowledge is the one that maximizes entropy subject to the known constraints. For a discrete set of possible outcomes $\{x_i\}$ with probabilities $p_i$, the Shannon entropy is defined as

$$H = -\sum_i p_i \log p_i$$

Suppose the information we have consists of expected-value constraints on a set of functions $f_k(x)$:

$$\sum_i p_i f_k(x_i) = F_k \text{ for all } k$$

The problem is to maximize $H$ under these constraints and the normalization condition $\sum_i p_i = 1$. Introducing Lagrange multipliers $\{\lambda_k\}$ and $\alpha$, we maximize

$$\Phi = -\sum_i p_i \log p_i - \alpha\left(\sum_i p_i - 1\right) - \sum_k \lambda_k \left(\sum_i p_i f_k(x_i) - F_k\right)$$

Setting derivatives to zero yields the canonical form of the maximum-entropy distribution:

$$p_i = \frac{1}{Z} \exp\left(-\sum_k \lambda_k f_k(x_i)\right)$$

where the partition function

$$Z = \sum_i \exp\left(-\sum_k \lambda_k f_k(x_i)\right)$$

ensures normalization. The Lagrange multipliers $\lambda_k$ are determined by the constraints $F_k$. This construction is formally identical to the Boltzmann–Gibbs distribution in statistical mechanics, confirming that statistical physics and rational inference share the same informational foundation.

## A.4 Relation Between Entropy and Bayesian Updating

In Bayesian inference, new evidence $E$ updates a prior $P(G)$ into a posterior $P(G \mid E)$. The information gain (or Kullback–Leibler divergence) associated with this update is

$$D_{KL}[P(G \mid E) \;\|\; P(G)] = \sum_G P(G \mid E) \log \frac{P(G \mid E)}{P(G)}$$

This quantity is always non-negative and equals zero only when the new evidence does not change belief. Hence, every act of learning corresponds to a reduction in entropy:

$$\Delta H = -D_{KL}[P(G \mid E) \;\|\; P(G)] \leq 0$$

The posterior distribution always contains less uncertainty than the prior, formalizing the intuitive notion that information reduces ignorance while preserving coherence.

## A.5 Induction as Consistency, Not Certainty

Finally, the Bayesian logic of plausible reasoning unifies induction and deduction under a single criterion of consistency. When information is complete, probabilities collapse to truth values:

$$P(G \mid E) \in \{0,1\}$$

When information is incomplete, they occupy the continuum between 0 and 1:

$$0 < P(G \mid E) < 1$$

In both cases, the same algebraic rules apply. Thus, deductive certainty and inductive plausibility are not distinct forms of reasoning but limiting cases of a single coherent system. The only difference lies in the degree of information available to the reasoner.

## Summary.

The mathematical framework presented here shows that the logic of plausible reasoning arises from fundamental principles of internal coherence and information conservation. Bayes's rule enforces consistency among beliefs; the Bayes Factor quantifies evidential strength; and the maximum entropy principle prescribes the most unbiased representation of ignorance. Together, they define a complete and unified foundation for rational inference — a solution to the problem of induction not by proof, but by logical necessity.

## Appendix B – Illustrative Computations

This appendix illustrates the numerical behavior of the models discussed in the main text through a series of figures rather than raw numerical tables. Each figure displays how posterior beliefs, evidential strength, and predictive probabilities evolve as data accumulate, thereby translating the logic of plausible reasoning into visual form. Throughout, let $E_1, \ldots, E_n$ be Bernoulli observations with $E_i = 1$ if the event occurs (e.g., "the Sun rises" on day $i$). Let $\theta \in (0,1)$ denote the Bernoulli success probability under model $M$.

### B.1 Laplace's model with a uniform prior

Assume a uniform prior on $\theta$, i.e. $\theta \sim \text{Beta}(1,1)$. If all $n$ observations are successes ($T_n = n$), the posterior is

$$f(\theta \mid T_n = n) = (n+1)\,\theta^n, \quad 0 < \theta < 1.$$

The predictive behavior implied by Laplace's uniform prior is shown in **Figure B1**, which plots the probability that the next event will be a success,

$$P(E_{n+1} = 1 \mid T_n = n) = \int_0^1 \theta\, f(\theta \mid T_n = n)\, d\theta = \frac{n+1}{n+2}.$$

as a function of the number of consecutive successes $n$.
The curve approaches one monotonically but never reaches it, illustrating that, under a continuous prior, even perfect evidence cannot justify a universal law ($P(\theta = 1 \mid T_n = n) = 0$).

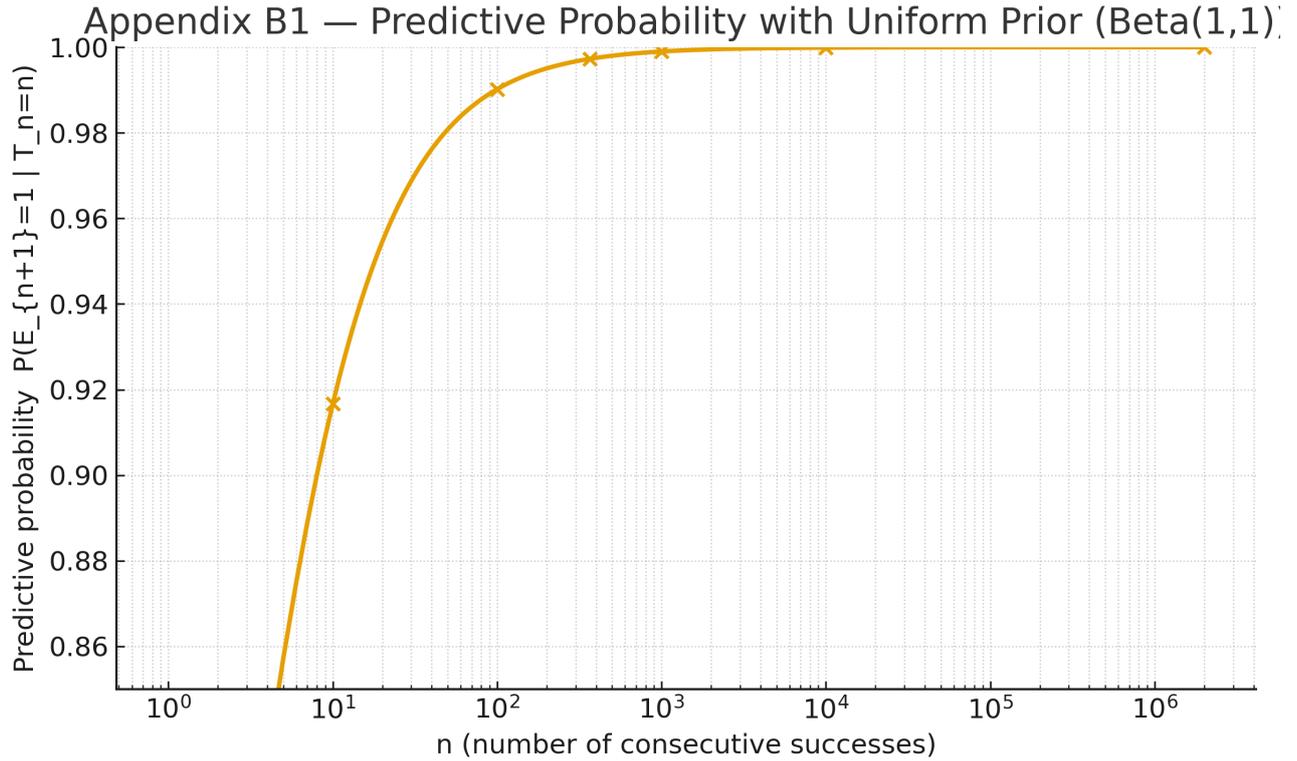

**Figure B1**. Predictive probability under Laplace's uniform prior Beta(1,1). The curve shows $P(E_{n+1} = 1 \mid T_n = n) = \frac{n+1}{n+2}$ as a function of the number of consecutive successes $n$ (log scale). The probability approaches one monotonically but never reaches it, illustrating that under a continuous prior even perfect evidence cannot justify a universal law ($P(\theta = 1 \mid T_n = n) = 0$).

## B.2 Jeffreys's mixed prior (point mass at the boundary)

To model the possibility of a universal law, Jeffreys assigns a prior with a **point mass at $\theta = 1$** and a continuous component on (0,1):

$$\pi(\theta) = \frac{1}{2}\delta(\theta - 1) + \frac{1}{2}U(0,1).$$

After observing $n$ consecutive successes ($T_n = n$), the posterior probability that the law holds (i.e., that $\theta = 1$) is

$$P(\theta = 1 \mid T_n = n) = \frac{\frac{1}{2} \cdot 1}{\frac{1}{2} \cdot 1 + \frac{1}{2} \cdot \frac{1}{n+1}} = \frac{n+1}{n+2}.$$

Hence the posterior mass on $\theta = 1$ **increases monotonically** with $n$ and approaches 1 as $n \to \infty$, though it never reaches certainty for finite $n$. The corresponding behavior is visualized in **Figure B2**, which plots $P(\theta = 1 \mid T_n = n)$ as a function of $n$ (log-scale) and highlights the benchmark values used in Appendix B1. (With prior weights other than 50–50, the same formula applies with $\frac{1}{2}$ replaced by $w \in (0,1)$; the qualitative behavior is unchanged.)

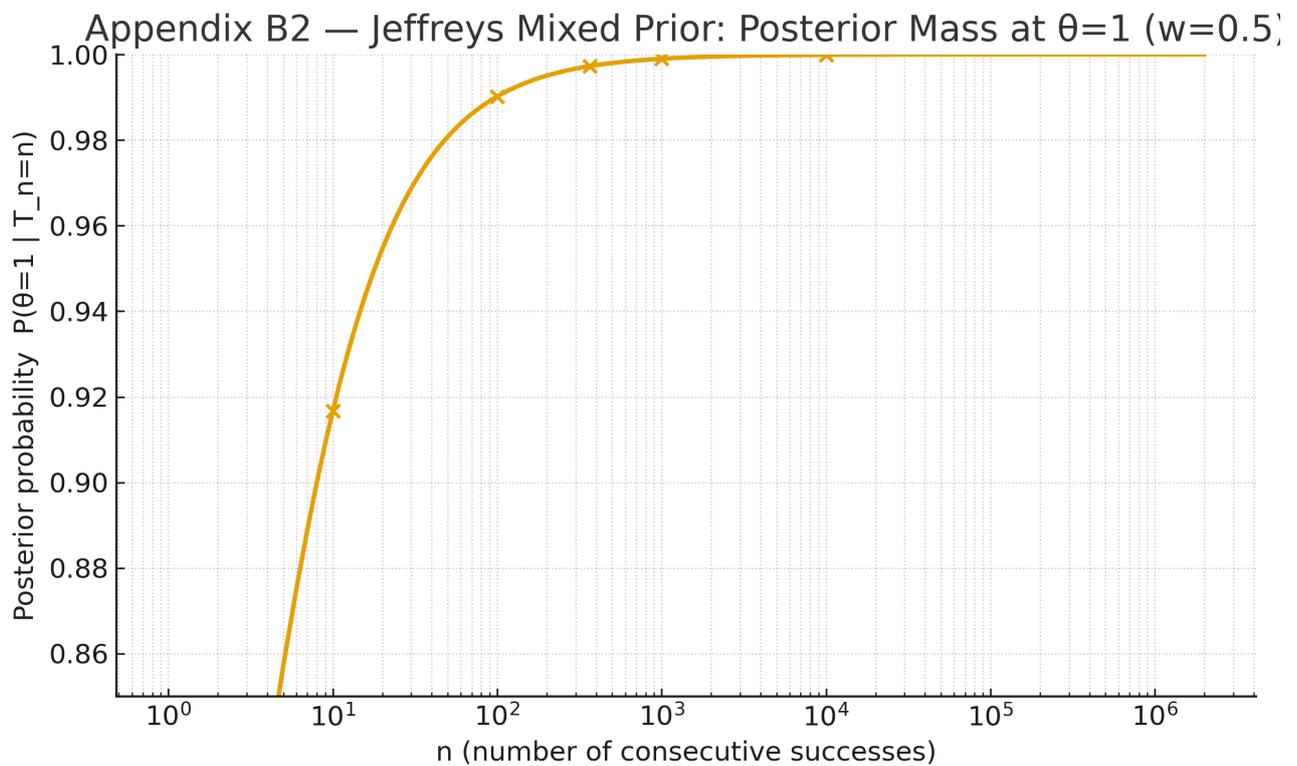

**Figure B2.** Jeffreys's mixed prior ($\pi = \frac{1}{2}\delta_1 + \frac{1}{2}U(0,1)$) yields a posterior point mass at $\theta = 1$ equal to $P(\theta = 1 \mid T_n = n) = \frac{n+1}{n+2}$. The curve (log-scale on $n$) shows monotonic convergence to 1 without ever reaching certainty for finite $n$; markers indicate the benchmark values used in Appendix B1.

## B.3 Bayes Factor for the sunrise law

To compare the universal hypothesis $G: \theta = 1$ against the composite alternative $G^C: \theta \sim \text{Beta}(1,1)$, the Bayes Factor after $T_n = n$ consecutive successes is

$$\text{BF}(G, G^C; T_n = n) = \frac{P(T_n = n \mid \theta = 1)}{P(T_n = n \mid \theta \sim \text{Beta}(1,1))} = \frac{1}{\int_0^1 \theta^n \, d\theta} = n + 1.$$

Thus, the evidence in favor of $G$ grows linearly with the number of perfect observations, producing a logarithmic increase on a decibel-like scale. The relationship between $n$ and $\log_{10} \text{BF}$ is displayed in **Figure B3**, which shows the monotonic accumulation of evidence toward the universal law.

If there is even a single failure among $n$ trials ($T_n = n - 1$), then
$$P(T_n = n - 1 \mid \theta = 1) = 0 \Rightarrow \text{BF}(G, G^C; T_n = n - 1) = 0,$$

corresponding to instantaneous evidential collapse for $G$.

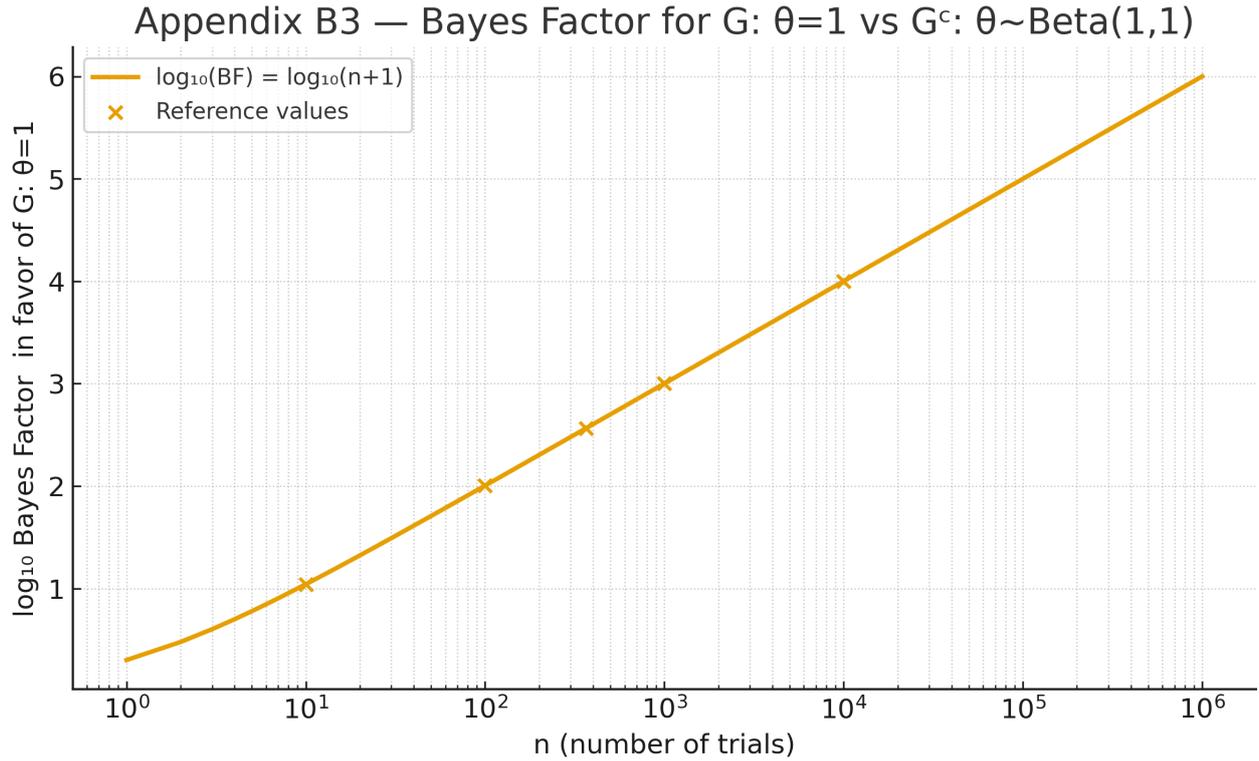

**Figure B3.** Bayes Factor in favor of the universal law $G: \theta = 1$ versus the composite model $G^C: \theta \sim \text{Beta}(1,1)$. The curve shows $\log_{10} \text{BF} = \log_{10}(n+1)$ as a function of $n$ (log scale). Evidence grows steadily with the number of perfect successes and collapses to zero after the first failure.

## B.4 One failure after many successes (update and prediction)

Suppose we observe $n-1$ successes and a single failure under a uniform prior. The posterior for the success probability is then

$$\theta \mid T_n = n - 1 \sim \text{Beta}(n, 2).$$

The predictive probability that the next observation will be a success is

$$P(E_{n+1} = 1 \mid T_n = n - 1) = \frac{n}{n+2}.$$

This value is slightly smaller than in the all-success case (Appendix B1), reflecting the rational adjustment after the first counterexample.

Figure B4 plots the predictive probability as a function of $n$ (log scale), showing that plausibility remains high for large $n$ but never returns to unity once a failure has occurred.

Under the universal hypothesis $G : \theta = 1$, the likelihood of any failure is zero; therefore, the Bayes Factor for $G$ collapses immediately to zero.

In a confidence-based framework, the acceptance indicator also flips from full to null after the first failure, demonstrating the on–off procedural behavior contrasted with the Bayesian's continuous reweighting of belief.

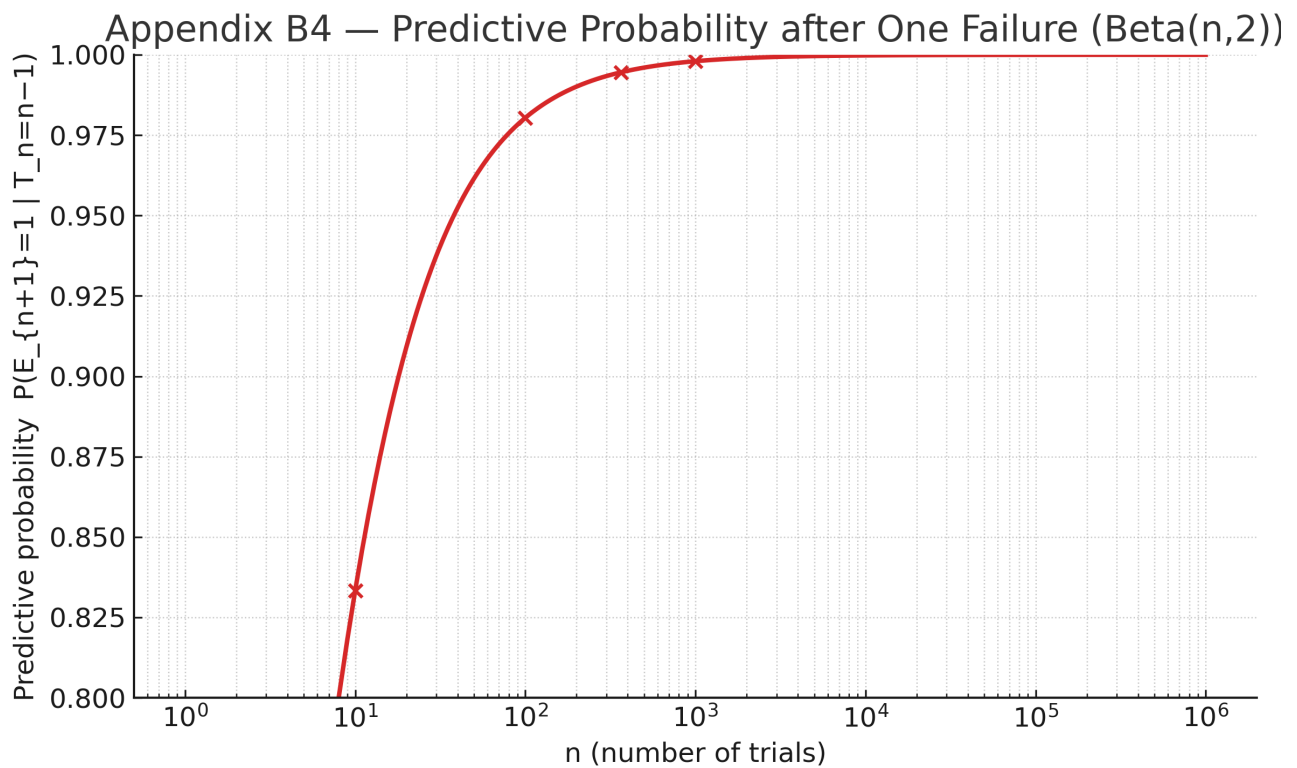

**Figure B4.** Predictive probability of success after one failure, $P(E_{n+1} = 1 \mid T_n = n - 1) = n/(n + 2)$, under a uniform prior. The curve (log scale on $n$) shows that plausibility decreases smoothly after a counterexample but remains high for large $n$, exemplifying the continuity of Bayesian updating compared with the binary behavior of confidence-based acceptance.

## B.5 Posterior credible interval (Beta–Binomial)

With a uniform prior and $t$ successes out of $n$, the posterior is
$$\theta \mid t, n \sim \text{Beta}(t+1, n-t+1).$$

A symmetric $(1-\alpha)$ credible interval is obtained from the Beta quantiles:
$$\text{CI}_{1-\alpha}(\theta \mid t, n) = [\, q_{\text{Beta}}(\alpha/2;\, t+1, n-t+1),\, q_{\text{Beta}}(1-\alpha/2;\, t+1, n-t+1)\,].$$

For the all-success case ($t = n$), this reduces to a $\text{Beta}(n+1, 1)$ posterior.

For large $n$, a normal approximation around the posterior mean
$$\mu = \frac{n+1}{n+2},\, \sigma^2 = \frac{\mu(1-\mu)}{n+3},$$

is often adequate:
$$\theta \mid T_n = n \approx \mathcal{N}\left(\frac{n+1}{n+2},\, \frac{((n+1)/(n+2))(1/(n+2))}{n+3}\right) \text{ (truncated to } [0,1]).$$

**Figure B5** illustrates the posterior mean and 95% credible interval for the all-success case, showing how the interval narrows rapidly as $n$ increases while the mean approaches one without ever reaching it.

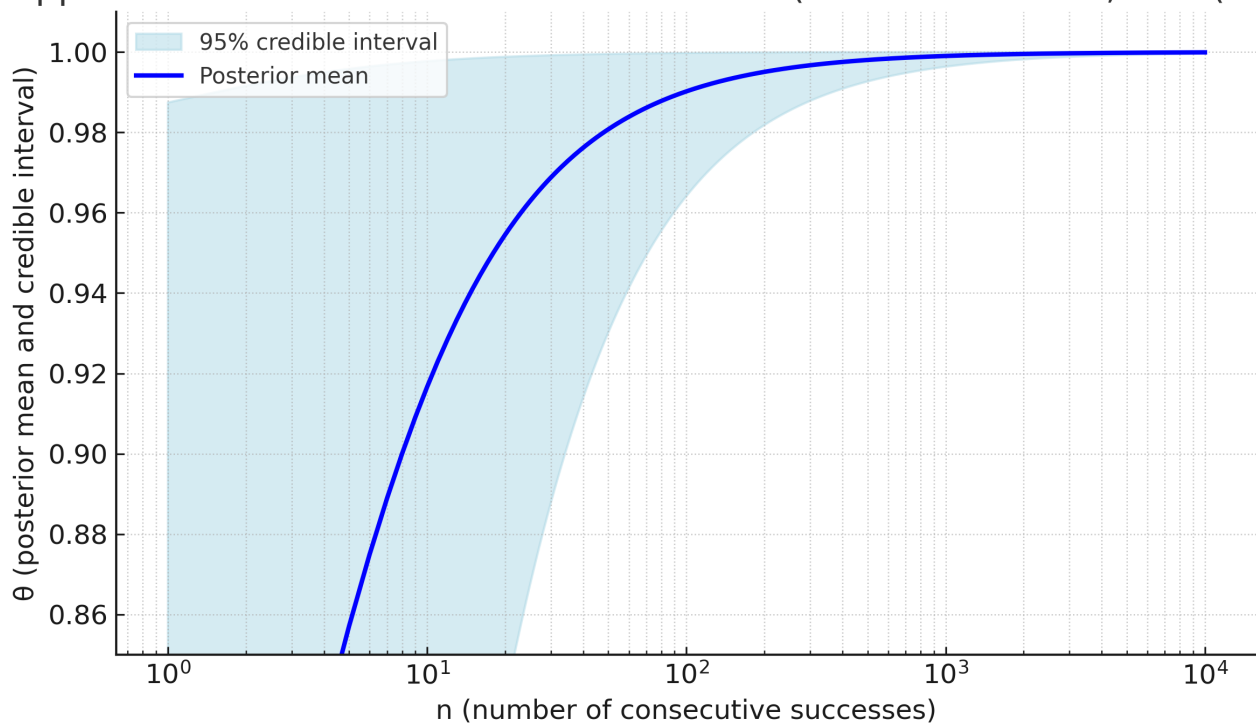

**Figure B5.** Posterior mean (solid line) and 95% credible interval (shaded area) for $\theta$ under a uniform prior and all-success data ($t = n$). As $n$ grows, the interval collapses around the mean $(n + 1)/(n + 2)$, approaching but never attaining certainty ($\theta = 1$).

### B.6 Summary table (at a glance)

- **Laplace (uniform prior):**

$$P(E_{n+1} = 1 \mid T_n = n) = \frac{n+1}{n+2}, P(\theta = 1 \mid T_n = n) = 0.$$

- **Jeffreys (mixed prior ½–½):**

$$P(\theta = 1 \mid T_n = n) = \frac{n+1}{n+2}.$$

- **Bayes Factor (law vs Beta–uniform):**

$$BF(G, G^C ; T_n = n) = n + 1, BF(G, G^C ; T_n < n) = 0.$$

- **After one failure (uniform prior)**:

$$\theta \mid T_n = n - 1 \sim \text{Beta}(n, 2), P(E_{n+1} = 1 \mid T_n = n - 1) = \frac{n}{n+2}.$$

These computations illustrate the core message: Bayesian updating provides a coherent, quantitative account of plausibility (prediction and learning), while *confidence* procedures supply acceptance rules with on–off behavior; neither transforms finite evidence into certain universal truth, but Bayes gives a principled measure of how evidence reshapes belief.

# Appendix C – Notation and Definitions

This appendix summarizes the main symbols and expressions used throughout the manuscript. All probabilities are conditional, expressing degrees of plausibility in the sense of Jaynes. Upper-case letters (e.g., $G, E, H$) denote propositions; lower-case Greek letters (e.g., $\theta, \lambda$) denote parameters or hyperparameters.

| Symbol (short) | Meaning | Domain / Notes |
|---|---|---|
| $P(A\|B)$ | Probability / plausibility of A given B | Values in [0, 1] |
| $L(\theta; E)$ | Likelihood of parameter θ given evidence E | Non-negative real |
| $\pi(\theta)$ | Prior for θ (before data) | Probability density |
| $f(\theta\|E)$ | Posterior for θ (after data) | Normalized density |
| $BF_{12}$ | Bayes Factor comparing $G_1$ and $G_2$ | Ratio of marginal likelihoods |
| $C(G; E)$ | Confidence measure for G under E | Range [0, 1] |
| $U(a\|E)$ | Expected utility of action a given E | Real number |
| $H$ | Shannon entropy (uncertainty measure) | $\geq 0$ |
| $D_{kl}[p\|\|q]$ | Kullback–Leibler divergence (info gain) | $\geq 0$; = 0 iff p = q |
| $Z$ | Partition / normalization constant | Positive real |
| $\delta(\cdot)$ | Dirac delta function | Unit mass at a point |
| $U(0,1)$ | Uniform distribution on [0, 1] | — |
| $\mathrm{Beta}(\alpha, \beta)$ | Beta distribution (shape α, β) | PDF on [0, 1] |
| $CI_{1-\alpha}$ | $(1 − \alpha)$ credible / confidence interval | Interval subset of parameter space |

**Key Equations**

**Product rule**

$$P(AB \mid C) = P(A \mid C)\, P(B \mid AC)$$

**Sum rule**

$$P(A + B \mid C) = P(A \mid C) + P(B \mid C) - P(AB \mid C)$$

**Bayes's theorem**

$$P(G \mid E) = \frac{P(E \mid G)\, P(G)}{P(E)}$$

**Bayes Factor**

$$BF_{12} = \frac{P(E \mid G_1)}{P(E \mid G_2)}$$

**Maximum-entropy distribution**

$$p_i = \frac{1}{Z} \exp\left(-\sum_k \lambda_k f_k(x_i)\right)$$

**Information gain (Kullback–Leibler divergence)**

$$D_{KL}[P(G \mid E) \;||\; P(G)] = \sum_{G} P(G \mid E) \log \frac{P(G \mid E)}{P(G)} \geq 0$$